\documentclass[aps,pra,twocolumn,amsmath,amsfonts,showpacs,amssymb,floatfix,nofootinbib,superscriptaddress]{revtex4}
\usepackage{graphicx,graphics,psfrag,bm,bbm,times,color}
\newcommand{\be}{\begin{eqnarray}}
\newcommand{\ee}{\end{eqnarray}}
\begin{document}
\title{Temperature effects on mixed-state geometric phase}
\author{A. T. Rezakhani}
\affiliation{Institute for Quantum Information
Science, University of Calgary, Alberta T2N 1N4, Canada}
\affiliation{Institute for Scientific
Interchange (ISI), Villa Gualino, Viale Settimio Severo 65, I-10133
Torino, Italy}
\author{P. Zanardi}
\affiliation{Institute for Scientific
Interchange (ISI), Villa Gualino, Viale Settimio Severo 65, I-10133
Torino, Italy}

\begin{abstract}
Geometric phase of an open quantum system that is interacting with a
thermal environment (bath) is studied through some simple examples.
The system is considered to be a simple spin-half particle which is
weakly coupled to the bath. It is seen that even in this regime the
geometric phase can vary with temperature. In addition, we also
consider the system under an adiabatically time-varying magnetic
field which is weakly coupled to the bath. An important feature of
this model is that it reveals existence of a temperature-scale in
which adiabaticity condition is preserved and beyond which the
geometric phase is varying quite rapidly with temperature. This
temperature is exactly the one in which the geometric phase
vanishes. This analysis has some implications in realistic
implementations of geometric quantum computation.
\end{abstract}
\date{\today}
 \maketitle
\section{Introduction}
Geometric phase (GP) of a quantum state is one of important concepts
in quantum theory. Indeed, Berry was the first who emphasized the
geometric nature of the phase acquired by an eigenstate of an
adiabatically-varying Hamiltonian in a closed loop in parameter
space \cite{berry}. Since its discovery, this concept has been a
subject of a vast investigation and generalization in many aspects
\cite{shapere,anandan}. An important reason for the interest in the
concept of GP is its relevance to geometric quantum computation
\cite{gqc}. Indeed, it is believed that the purely geometric
characteristic of such phases potentially provides robustness
against certain sources of noise \cite{rob}.

For general evolutions of an open quantum system, where the dynamics
is generally nonunitary, the GP has been defined in different
methods
\cite{uhl1,gam,ericsson2,pix,pati,carollo1,carollo2,nonunitary,lidar,sanders}.
The first general approach is Uhlmann's mathematical method which is
based on a purification of mixed states \cite{uhl1}. Another, more
recent, approach is based on a kinematic extension of the
interferometric approach firstly used in the case of unitary
evolutions \cite{nonunitary}. These methods are generally argued to
be different \cite{difference}, however, recently a formal approach
bridging between them has also been proposed \cite{ours}.

 As mentioned above studying how GP is affected by environmental noises is an important issue in
investigation of robustness of geometric quantum computation. In
some earlier works, effect of different types of decoherence sources
on GP has been studied.
In Ref.~\cite{carollo2}, it has been shown that the GP of a
spin-half system driven by one or two mode quantum fields subject to
decoherence behaves differently for adiabatic and nonadiabatic
evolutions. Also in Ref.~\cite{Yiwang1}, the GP of a two-level
system driven by a quantized magnetic field subject to dephasing has
been considered and it has been shown that the GP acquired by the
system can be observed even in long time-scales. Ref.~\cite{sanders}
addresses the change of the mean GP when the two-level system is
weakly coupled to a thermal. An interesting feature of this
investigation is that the mean GP does not have any thermal
correction up to the first order in the coupling constant, that is,
any dependence on temperature is of higher orders in the coupling
constant. In Ref.~\cite{whitney1}, it has been shown that coupling
to an environment induces some geometric and nongeometric
corrections to the Berry phase (BP) of a spin-half system that is
under an adiabatically slow rotating magnetic field. Moreover, it
has been argued that the BP can be observed only in experiments
whose time-scales, on one hand, are slow enough to ignore
nonadiabatic correction, and fast enough, on the other hand, to
prevent dephasing from deleting all phase information. Existence of
such a finite (adiabaticity) time-scale for the GP, indeed, has been
shown to be a general feature of open quantum systems
\cite{lidar,sarandy1}. Some other related studies about decoherence
effects on the GP can be found in
Refs.~\cite{romero,dechiaraPRL,dec,tidstrom}.

\section{The models}
In this paper we consider a simple spin-half particle in a magnetic
field $\vec{B}(t)$ which is weakly coupled to an environment (a
thermal bath, for example, consisting of photons) with temperature
$T$. We are mainly interested in finding thermal effects of the
environment on the GP of the system. To be specific, here, we follow
the kinematic definition of the GP ~\cite{nonunitary}. If the
density matrix of the system evolves as
$\varrho(0)\rightarrow\varrho(t)=\sum_ip_i(t)|\textit{w}_i(t)\rangle\langle
\textit{w}_i(t)|$, where $t\in[0,\tau]$ and $p_i(t)$'s
($|\textit{w}_i(t)\rangle$'s) are eigenvalues (eigenvectors) of
$\varrho(t)$, then the GP acquired by this state during this
interval reads as follows: \be\label{tongformula} \hskip
-4mm&\Phi=\text{arg}\sum_{i}[p_i(0)p_i(\tau)]^{\frac{1}{2}}\langle
\textit{w}_i(0)|\textit{w}_i(\tau)\rangle e^{-\int_0^{\tau}\langle
\textit{w}_i(t)|\dot{\textit{w}}_i(t)\rangle\text{d}t}.\hskip 1mm\ee
Throughout this paper we always consider the GP measured in
$\mod2\pi$. To use the above relation one must solve the eigenvalue
problem for the density matrix of the system to find eigenvalues as
well as the corresponding eigenvectors.

\begin{figure}[tp]
\includegraphics[width=6.50cm,height=3.8cm]{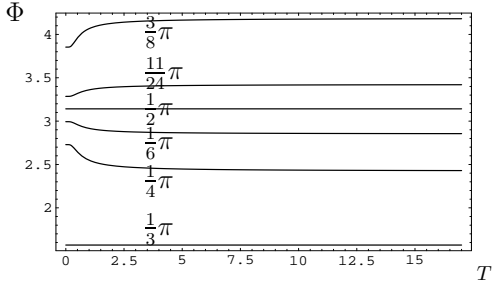}
\vskip -3mm\caption{GP ($\Phi$, in radian) vs. temperature ($T$, in
the units where $\hbar\equiv k_B\equiv 1$) for different values of
$\vartheta$. Here, we have taken $\epsilon=1/3$ and $\Delta=1$ (in
the same natural units).} \label{fig00}
\end{figure}

 Before discussion about coupling of an environment and the system in
 a relatively realistic model, let us illustrate a couple of simple examples that can reveal
 some general features of more complicated models. As the first example, suppose
that due to interaction with the environment, our system which is a
spin-half particle, initially is prepared in the thermally impure
state $(1-\epsilon)|+;0\rangle_n\langle+;0|+\epsilon
\varrho_{\text{th.}}$, where $|+;t\rangle_n
=\left(\begin{smallmatrix}\cos\frac{\vartheta(t)}{2}e^{-\frac{i}{2}\varphi(t)}\\
\sin\frac{\vartheta(t)}{2}e^{\frac{i}{2}\varphi(t)}\end{smallmatrix}\right)$
is an eigenvector of $\textbf{n}.\vec{\sigma}$,
$\textbf{n}=(\sin\vartheta\cos\varphi,\sin\vartheta\cos\varphi,\cos\vartheta)$,
$\varrho_{\text{th.}}=\lambda|+;0\rangle_n\langle+;0|+(1-\lambda)
|-;0\rangle_n\langle-;0|$, $\frac{\lambda}{1-\lambda}=e^{-\Delta/T}$
(we work in the natural units where $\hbar\equiv k_B\equiv 1$),
$\Delta$ is the energy gap of the system, $\vec{\sigma}=(\sigma_x,\sigma_y,\sigma_z)$ is the vector of the Pauli matrices, and for simplicity we take
$\vartheta(t)=\vartheta(0)$ and $\varphi(t)=\omega_0t$. Now let us
assume that the density matrix of the system has the following
(unitary) evolution during the period
$t\in[0,\frac{2\pi}{\omega_0}]$ \be&[(1-\epsilon)+\lambda\epsilon]
|+;t\rangle_n\langle+;t|+(1-\lambda)\epsilon
|-;t\rangle_n\langle-;t|. \ee By using the general formula
(\ref{tongformula}) one can easily obtain the GP acquired by this
state after its cyclic evolution as follows: \be &\Phi=\pi+\arctan
\left(\tan(\pi\cos\vartheta)[1-2\epsilon/(1+e^{-\Delta/T})]\right).
\ee Figure~\ref{fig00} shows that depending on the value of
$\vartheta$ the GP may decrease or increase with temperature. We
will see a bit later that the GP shows such a typical behavior even
in more complicated cases.

As the second example, let us consider two coupled spin-half
particles, one of which driven by a time-varying magnetic field,
with the Hamiltonian \be &H(t)=\frac{1}{2}\vec{B}(t).\vec{\sigma}_1+J(\sigma_{+1}\sigma_{-2}+\text{H.c.}),\ee
where $\sigma_{\pm}=\frac{1}{2}(\sigma_x\pm i\sigma_y)$ and the indices 1 and 2 indicate the corresponding
subsystems. The magnetic field is considered to be $B\textbf{n}(t)$,
where $\textbf{n}(t)$ has a form as in the previous example but it
rotates about the $z$-axis adiabatically. In fact, in this simple
case the second particle can be taken as an environment which is
coupled to our system of interest. The eigenvalues ($E_j$) and
instantaneous eigenvectors ($|\Psi_j(t)\rangle$) of this Hamiltonian
can be found simply as
follows:\be
&E_{1(3)}=-E_{2(4)}=\sqrt{1+\frac{\kappa^2}{2}\pm\frac{\kappa}{2}\sqrt{\kappa^2+4\sin^2\vartheta}},
\\&|\Psi_j\rangle=\frac{1}{\sqrt{N_j}}(a_j|\uparrow\hskip-.7mm\downarrow\rangle+b_j|\upuparrows\rangle+c_j|\downdownarrows\rangle
+d_j|\downarrow\hskip-.7mm\uparrow\rangle), \ee where
$d_j=\frac{\kappa\sin\vartheta(\cos\vartheta-E_j)}{E_j^2-\cos
2\vartheta}e^{i\omega_0t}$,
$b_j=-\frac{E_j+\cos\vartheta}{\sin\vartheta}d_je^{-i\omega_0t}$,
$a_j=\sin\vartheta e^{-i\omega_0t}$, $c_j=E_j-\cos\vartheta$,
$N_j=|a_j|^2+b_j^2+c_j^2+|d_j|^2$, and $\kappa=\frac{2J}{B}$. In
Ref.~\cite{yiwang}, the BP's of the subsystems for all eigenvectors
have been calculated, and it has been shown that the BP of the
composite system is sum of the BP's of the two constituting
subsystems. Here, we suppose that the composite system is in the
thermally diluted state
$\rho_j=(1-\epsilon)|\Psi_j\rangle\langle\Psi_j|+\epsilon\rho_{\text{thermal}}$,
in which $\rho_{\text{thermal}}\propto e^{-H/T}$. We are interested
in calculating the BP of the subsystem 1 to see how it is affected
by this simple thermalization mechanism. By a simple calculation, it
can be seen that in the Schmidt decomposition of the eigenvectors
coefficients are time-independent,
$|\Psi_j(t)\rangle=\sum_{\alpha}\sqrt{p_{\alpha}^j}|F_{\alpha}^j(t)\rangle_1|f_{\alpha}^j(t)\rangle_2$.
This fact in turn results into the reduced density matrix
$\rho_1^j(t)=\text{tr}_2(\rho_j)=\sum_{i\alpha}\tilde{p}_{\alpha}^i|F_{\alpha}^i(t)\rangle\langle
F_{\alpha}^i(t)|$ for the subsystem 1, in which for $i=j$ we have
$\tilde{p}_{\alpha}^j=[1-\epsilon+\epsilon
e^{-E_j/T}/\text{tr}(e^{-H/T})]p_{\alpha}^j$ and
$\tilde{p}_{\alpha}^i=\epsilon
e^{-E_i/T}p_{\alpha}^i/\text{tr}(e^{-H/T})$ otherwise. To calculate
the BP of this density matrix we use the idea of purification
\be&\Phi_{\text{Berry}}(\rho_1^j(t))=\Phi_{\text{Berry}}(|\Upsilon^j(t)\rangle_{1a}),\ee
where
$|\Upsilon^j(t)\rangle_{1a}=\sum_{i\alpha}\sqrt{\tilde{p}_{\alpha}^i}|F_{\alpha}^i(t)\rangle_1|a_{\alpha}^i\rangle_a$
is a purified state obtained from $\rho_1^j(t)$ by attaching an
ancilla $a$. By using this purification and considering constancy of
the coefficients of the Schmidt decompositions, it can easily be
\begin{figure}[tp]
\includegraphics[width=8cm,height=5.2cm]{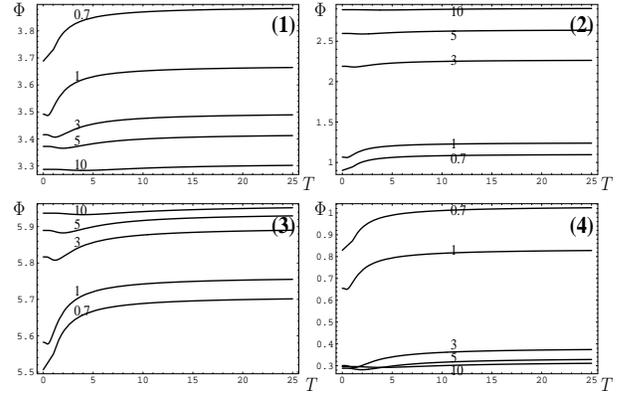}
\vskip -1mm\caption{BP ($\Phi$, in radian) of the subsystem 1 vs.
temperature ($T$, in the units where $\hbar\equiv k_B\equiv 1$),
obtained from $\rho_1^j$, $j=1,2,3,4$, for $\vartheta=\pi/4$, with
$\epsilon=0.1$. The numbers show the related values of the coupling
constant $\kappa$.} \label{figberry}
\end{figure}
\begin{figure}[bp]
\includegraphics[width=6.5cm,height=3.8cm]{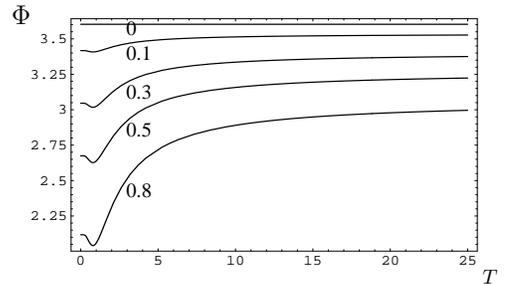}
\vskip -3mm \caption{BP ($\Phi$, in radian), vs. temperature ($T$,
in the units where $\hbar\equiv k_B\equiv 1$) for different
values of $\epsilon$. Here, we have taken $\rho_1^1$, and
$\kappa=2$, $\vartheta=\pi/4$.} \label{figbb}
\end{figure}
concluded that
\be&\Phi_{\text{Berry}}(\rho_1^j)=\sum_{i\alpha}\tilde{p}_{\alpha}^i\Phi_{\text{Berry}}(F_{\alpha}^i),\ee
where
$\Phi_{\text{Berry}}(F_{\alpha}^i)=i\int_0^{\frac{2\pi}{\omega_0}}\langle
F_{\alpha}^i(t)|\dot{F}_{\alpha}^i(t)\rangle\text{d}t$.
Figure~\ref{figberry} illustrates the BP of the subsystem 1 during a
cyclic evolution, for $j=1,2,3,4$. As is seen, after a decrease for
some temperature ranges, when temperature increases the BP increases
as well. Figure~\ref{figbb} shows that the more thermal the
composite system is the lower the value of the BP is. However, even
in this case the general behavior of the BP vs. temperature is as in
Fig.~\ref{figberry}.

Now, let us consider a more realistic case in which the spin is
interacting with a thermal environment consisting of photons. It is
known that when the spin is not isolated its dynamics is much more
involved. When the magnetic field is not time-varying, interaction
with the environment generally induces energy and phase relaxation
processes (with the time-scales $T_1$ and $T_2$, respectively), in
addition to a Lamb shift of the energy levels ($\delta
E_{\text{Lamb}}$) \cite{alickilendi}. It is natural to expect that
when dephasing is the leading regime of the interaction no GP effect
could be observed \cite{whitney1}. However, in the weak-coupling
limit, defined by $B\gg T_2^{-1}$, in experiment time-scale $\tau$
these effects can be seen if both adiabaticity condition
$B\gg\tau^{-1}$ and no-dephasing regime condition $\tau\lesssim T_2$
can be satisfied simultaneously \cite{whitney1}. To fulfill these
conditions, in our discussion we assume that the field varies
adiabatically on a closed loop in the parameter space in time
$\tau$. Additionally, we also suppose that the time-scale in which
typical correlation functions of the environment decay,
$t_{\text{env.}}$, is much smaller than $\tau$ and all
dissipation-induced time-scales, i.e., $t_{\text{env.}}\ll\tau,
T_{\text{diss.}}=\min\{T_1,T_2,\delta E_{\text{Lamb}}^{-1}\}$. These
assumptions overall enable us to use Markovian approximation in the
form of a Lindblad-type master equation \cite{lindblad}
\be\label{lindbladeq}
&\dot{\varrho}=-i[H,\varrho]+\frac{1}{2}\sum_k\left([\Gamma_k,\varrho\Gamma^{\dagger}_k]+
[\Gamma_k\varrho,\Gamma^{\dagger}_k]\right), \ee where $H$ is the
effective Hamiltonian of the (open) system and $\Gamma_k$'s are the
quantum jump operators describing system-environment coupling.
\begin{figure}[tp]
\includegraphics[width=8.6cm,height=3cm]{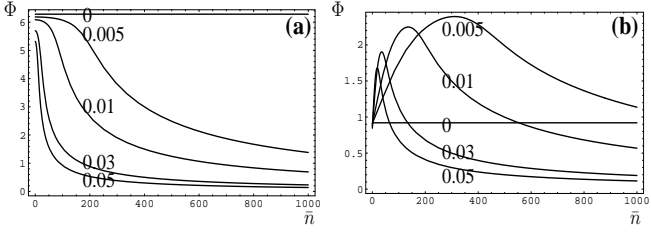}
\caption{GP ($\Phi$, in radian), vs. mean number of photons ($\bar{n}$) in the environment. 
Here, $\omega_0=2$, $\vartheta=\frac{\pi}{2}$ (a),
$\frac{\pi}{2}+\frac{\pi}{4}$ (b), and the numbers associated to
each curve denote the related value of the coupling constant
$\kappa$.} \label{fig6}
\end{figure}
The magnetic field is supposed to be in the direction
$\textbf{n}(t)=(\sin\vartheta\cos\varphi(t),\sin\vartheta\sin\varphi(t),\cos\vartheta)$,
where $\varphi(t)=\omega_0t$, and $\tau=\frac{2\pi}{\omega_0}$ is
the time necessary for completion of a loop. The Hamiltonian of the
system is $H(t)=\frac{1}{2}\vec{B}(t).\vec{\sigma}$, where
$\sigma_z=|e\rangle\langle e|-|g\rangle\langle
g|\equiv\left(\begin{smallmatrix}1&0\\0&-1\end{smallmatrix}\right)$,
in which $|g\rangle$ ($|e\rangle$) is the ground (excited) state of
the system in the absence of external magnetic field. For such an
open system the quantum jump operators are
$\Gamma_1=\sqrt{\kappa(\bar{n}+1)}|g\rangle\langle e|$, which
describes spontaneous and stimulated emissions into the bath, and
$\Gamma_2=\sqrt{\kappa \bar{n}}|e\rangle\langle g|$, which is the
corresponding absorption from the bath, where
$\bar{n}(\omega_0,T)=(e^{\omega_0/T}-1)^{-1}$ is the mean number of
photons in the bath. A master equation in the Lindblad form usually
can be applied in the case of time-independent Hamiltonians under
Markovian conditions. The simpler case in which the Hamiltonian of
the system is $\frac{1}{2}\omega_0\sigma_z$, a constant magnetic
field along $z$-axis, has been studied earlier in
Ref.~\cite{sanders}, and to the first order in
coupling constant $\kappa$, the mean GP has been calculated. 
It has been shown that up to this order no thermal correction
appears in the GP, that is, any dependence on temperature (through
$\bar{n}$) is of higher orders of $\kappa$. Figure \ref{fig6}
depicts temperature-dependence of the GP for the two values of
$\vartheta=\pi/2,3\pi/4$ and different values of the coupling
constant $\kappa$. For the values $\vartheta<\frac{\pi}{2}$ the case
is similar to $\vartheta=\frac{\pi}{2}$. When
$\vartheta\leqslant\frac{\pi}{2}$, this figure shows a decrease in
the GP with temperature, whereas for $\vartheta>\frac{\pi}{2}$ the
GP can also show an increase for some temperature range whose size
decreases when $\kappa$ increases.

Now, let us consider the general case with a time-varying magnetic
field, as explained above. To be able to use the Lindblad master
equation we change our reference frame and go to the diagonal frame
in which the Hamiltonian has the simple time-independent form
$h\sigma_z$, where
$h=\frac{1}{2}B\sqrt{\sin^2\vartheta+(\cos\vartheta-\omega_0/B)^2}$
\cite{romero}. Here it is worth noting that another alternative
method to consider the problem is to define an adiabatic Hamiltonian
\cite{vidal}, which in the absence of interaction with environment
shows the BP effect. In this way, in contrast to the rotating frame
method, only the length of magnetic field changes. The results of
this method have been shown to be equivalent to the rotating frame
method in the first order of the rotation frequency \cite{dec}. The
density matrix of the system in the diagonal frame,
\begin{figure}[tp]
\includegraphics[width=8cm,height=2.1cm]{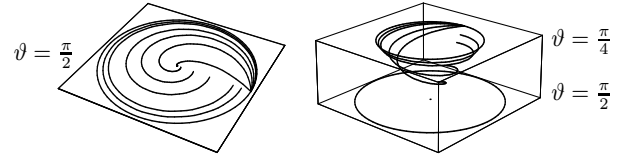}
\caption{The path traversed by the tip of the Bloch vector of the
spin starting from $\vartheta=\frac{\pi}{2}$, for
$\bar{n}=0,1,5,10,20,80$ (left), and $\vartheta=\pi/4$, for
$\bar{n}=0,1,5,10,20$ (right). In both of the figures, we have set
$\kappa=10^{-2}$, $B=10^3$, and $\omega_0=2$.} \label{fig4}
\end{figure}
\begin{figure}[bp]
\includegraphics[width=6.50cm,height=3.6cm]{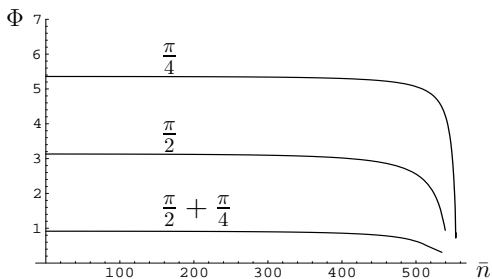}
\caption{GP ($\Phi$, in radian) vs. mean number of photons
($\bar{n}$) in the environment. Here, $\omega_0=2$,
$\kappa=10^{-3}$, $B=10^3$, and the azimuthal angles
$\vartheta=\frac{\pi}{4},\frac{\pi}{2},\frac{\pi}{2}+\frac{\pi}{4}$
are chosen for this plot.} \label{fig1}
\end{figure}
denoted by $\tilde{\varrho}$, is related to the density matrix in
the lab frame as
$\varrho(t)=e^{-\frac{i}{2}t\omega_0\sigma_z}V\tilde{\varrho}(t)Ve^{\frac{i}{2}t\omega_0\sigma_z}$,
where $V=\alpha_-\sigma_x+\alpha_+\sigma_z$, and
$\alpha_{\pm}=\sqrt{\frac{1}{2}\pm\frac{1}{4h}(B\cos\vartheta-\omega_0)}$.
The explicit form of the master equation in this frame is as follows:
\be \aligned
\dot{\tilde{\varrho}}=&{-i}[h\sigma_z,\tilde{\varrho}]+\kappa(\bar{n}+1)(2\sigma_-\tilde{\varrho}\sigma_+
-\tilde{\varrho}\sigma_+\sigma_--\sigma_+\sigma_-\tilde{\varrho})\\&+\kappa\bar{n}(2\sigma_+\tilde{\varrho}\sigma_-
-\tilde{\varrho}\sigma_-\sigma_+-\sigma_-\sigma_+\tilde{\varrho}).
\endaligned\ee
For simplicity we have omitted the Lamb shift terms, because the
Lamb shift contributes to the GP for $\tau\neq 2\pi/{\omega_0}$ only
\cite{sanders}. These equations can be solved exactly as follows:
\be\label{sol} \aligned
\tilde{\varrho}_{11}(t)&=\tilde{\varrho}_{11}(0)e^{-2\kappa
t(2\bar{n}+1)}+\frac{\bar{n}}{1+\bar{n}}[1-e^{-2\kappa
t(2\bar{n}+1)}]\\
\tilde{\varrho}_{12}(t)&=\tilde{\varrho}_{12}(0)e^{(-2ih+\kappa(2\bar{n}+1))t},
\endaligned\ee
where $\tilde{\varrho}_{11}$ ($\tilde{\varrho}_{12}$) is the
diagonal (off-diagonal) element of $\tilde{\varrho}$. We assume that
the initial state of the spin is
$|\Psi_0\rangle=\cos\frac{\vartheta}{2}|e\rangle+\sin\frac{\vartheta}{2}|g\rangle$.
Unlike the case of unitary evolution of mixed states where the GP
can be explained only by length of the Bloch vector ($r$) and the
solid angle subtended by it on the Bloch sphere ($\Omega$), i.e.
$\Phi=-\arctan\left(r\tan\frac{\Omega}{2}\right)$, in the case of
nonunitary evolution such information is not sufficient and the GP
has a complicated dependence on the Bloch vector. However, dynamics
of the Bloch vector can give an insight about how dissipative
mechanisms affect the system. Figure \ref{fig4} illustrates these
effects on the Bloch vector of the density matrix. As is seen, in
this case the length of the Bloch vector decreases, which
characterizes a coherence loss, and, the vector precesses toward the
center of the sphere, which is related to dissipation. After coming
back to the lab frame, to find the GP of the state one has to solve
the eigenvalue problem of $\varrho(t)$, and plug the result into
Eq.~(\ref{tongformula}). The explicit form of the GP is very
tedious, instead, we plot the results of some special cases in
Fig.~(\ref{fig1}). It shows that for a relatively large temperature
range the GP is varying very slowly with temperature. This is in
accordance with the results of Ref.~\cite{sanders}. In addition,
Fig.~\ref{mainfig}(a) shows that when temperature gradually
increases the GP decreases so that in some threshold value it
vanishes.
\begin{figure}[tp]
\includegraphics[height=2.9cm,width=8.5cm]{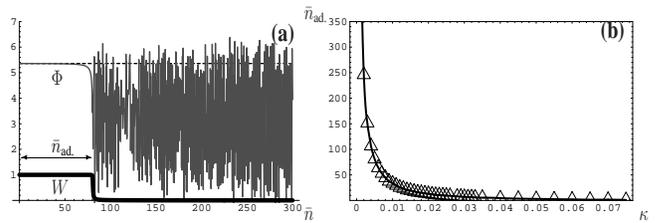}
\vskip -3mm \caption{(a) GP ($\Phi$, in radian) vs. mean number of
photons ($\bar{n}$) in environment (thin curve) and $W=|\langle
\textit{w}_1(0)|\textit{w}_1(\tau)\rangle|$ (thick curve). Here,
$\omega_0=2$, $B=10^3$, $\kappa=5\times 10^{-3}$,
$\vartheta=\frac{\pi}{4}$. (b) $\bar{n}_{\text{ad.}}$ vs. coupling
constant $\kappa$. Here, we have set $\omega_0=2$ and $B=10^3$.
}\label{mainfig}
\end{figure}
After this value the GP shows a rapidly varying behavior. The value
of this threshold depends on the parameters of the problem. An
special and important feature of this figure is that the overlap of
$|\Psi_0\rangle\equiv |\textit{w}_1(0)\rangle$ and
$|\textit{w}_1(\tau)\rangle$, $W=|\langle
\textit{w}_1(0)|\textit{w}_1(\tau)\rangle|$, as a measure of
adiabaticity of the evolution vanishes in the same threshold
temperature. This indicates that after this temperature adiabaticity
condition breaks down. This implies existence of a temperature-scale
$\bar{n}_{\text{ad.}}$ for the observation of the adiabatic GP. This
is similar to the existence of a finite adiabaticity time-scale for
observation of the GP, as pointed out in Ref.~\cite{lidar}. In fact,
this similarity between temperature and time can be understood
easily by looking back at Eq.~(\ref{sol}). Figure~\ref{mainfig}(b)
shows this threshold temperature vs. the coupling constant. As
expected, when the coupling constant increases the adiabaticity
temperature-scale decreases.

\section{ Conclusion}
We have studied thermal effects induced by an environment on the
geometric phase. It has been shown that even in weak-coupling and
adiabatic limits geometric phase can vary with the temperature of
the environment. One of implications of the results of this paper is
that in geometric phase experiments in systems which are in contact
to a dissipative environment care must be taken in correct
interpretation of results. The existence of an adiabaticity
temperature-scale, similar to the adiabaticity time-scale, puts some
constraints on the realization schemes of geometric quantum
information processing. This in turn can have some implications in
the more realistic investigations of robustness of geometric quantum
computation \cite{wuzanardilidar}.

\begin{acknowledgments}
This work has been supported by EU project TOPQIP under Contract No.
IST-2001-39215. A.T.R also acknowledges support by {\emph{i}}CORE and PIMS fellowships.
\end{acknowledgments}


\end{document}